\documentclass{article}
\usepackage{spconf,amsmath,graphicx}

\usepackage{enumitem}
\setlist{nosep, leftmargin=14pt}

\usepackage{mwe} % to get dummy images
\usepackage{float}

\title{GPU-Net: Lightweight U-Net with More Diverse Features}
\name{Heng Yu$^{\star \dagger}$ \qquad Di Fan$^{\star}$ \qquad Weihu Song$^{\dagger}$}

\address{$^{\star \dagger}$ Tsinghua University \\
    $^{\star}$ University of Southern California \\
    $^{\dagger}$ Hebei University of Engineering}
\begin{document}
%\ninept
%
\maketitle
\begin{abstract}
Image segmentation is an important task in the medical image field and many convolutional neural networks (CNNs) based methods have been proposed, among which U-Net and its variants show promising performance. In this paper, we propose GP-module and GPU-Net based on U-Net, which can learn more diverse features by introducing Ghost module and atrous spatial pyramid pooling (ASPP). Our method achieves better performance with more than $4\times$ fewer parameters and $2\times$ fewer FLOPs, which provides a new potential direction for future research. Our plug-and-play module can also be applied to existing segmentation methods to further improve their performance.
\end{abstract}
\section{Introduction}
\label{sec:intro}

Recently, many impressive CNN-based image segmentation models have been proposed. However, it is hard to apply these natural image segmentation models into medical image segmentation directly since there exist domain gaps. Aiming at this, researchers proposed U-Net~\cite{ronneberger2015u} and U-Net based variants, which achieve remarkable performance. These variants mainly focus on optimization of network structure, which introduce more parameters and generate useful feature maps inefficiently. In this paper, we explore the enhancement that well-learned features can bring and boost U-Net with more higher quality diverse features and fewer parameters. Our main contribution is proposing a lightweight version of U-Net which can achieve competitive and even better performance while significantly reduces the number of parameters and FLOPS. We name our method GPU-Net and experiments show GPU-Net can achieve state-of-the-art performance. To the best of our knowledge, it is the first paper that explore the possibility of applying ghost-module and its variants into U-Net. We believe our method can be an interest topic to discuss for the medical image segmentation community.

\section{Related Work}
\label{sec:relatedwork}

Image segmentation requires classifying each image pixel on an individual basis. While the deep learning research in natural image segmentation is in full swing, U-Net~\cite{ronneberger2015u} is proposed specifically for medical image segmentation. U-Net can be trained with a relatively small number of medical data from scratch and achieve competitive performance. Based on U-Net, a series of methods have been proposed. R2U-Net~\cite{alom2018recurrent} and Attention U-Net~\cite{oktay2018attention} apply recurrent module and attention mechanism into U-Net, respectively. These variants all develop the potential of U-Net from different angles but ignore the problem of feature redundancy. Redundancy in feature maps can be important but it is better to achieve this kind of redundancy using more efficient ways like getting a set of intrinsic feature maps first and then generating many ghost feature maps based on them as proposed in GhostNet~\cite{han2020ghostnet}. The intrinsic feature maps can have no redundancy so as to ensure efficiency. The increase in parameters also impair the innovation of architectures in their methods. In this paper, our target is to explore how U-Net can benefit from sufficient diverse features. To get more high quality features with simple operations in network, we borrow and improve the ghost module in GhostNet~\cite{han2020ghostnet} and propose GP-module, a lightweight module that can boost U-Net with more valuable features and fewer parameters. 

\begin{figure}[htb]
\begin{minipage}[b]{1.0\linewidth}
  \centering
  \centerline{\includegraphics[width=0.25\textwidth]{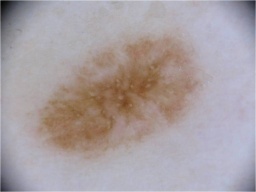}}
%  \vspace{2.0cm}
  \centerline{(a) Image}\medskip
\end{minipage}
\begin{minipage}[b]{.45\linewidth}
  \centering
  \centerline{\includegraphics[width=0.95\textwidth]{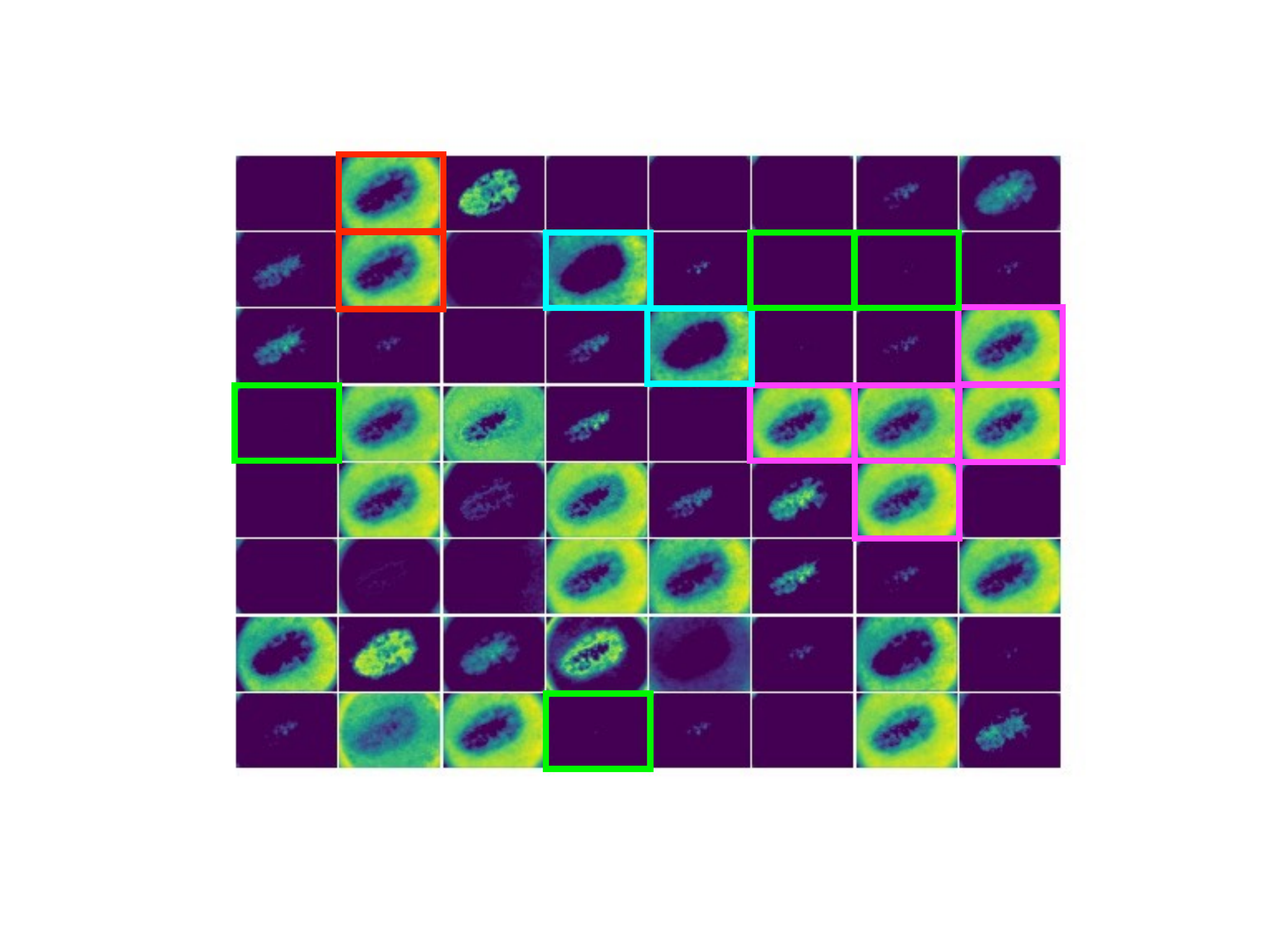}}
%  \vspace{1.5cm}
  \centerline{(b) First Convolution Feature}\medskip
\end{minipage}
\hfill
\begin{minipage}[b]{0.45\linewidth}
  \centering
  \centerline{\includegraphics[width=0.95\textwidth]{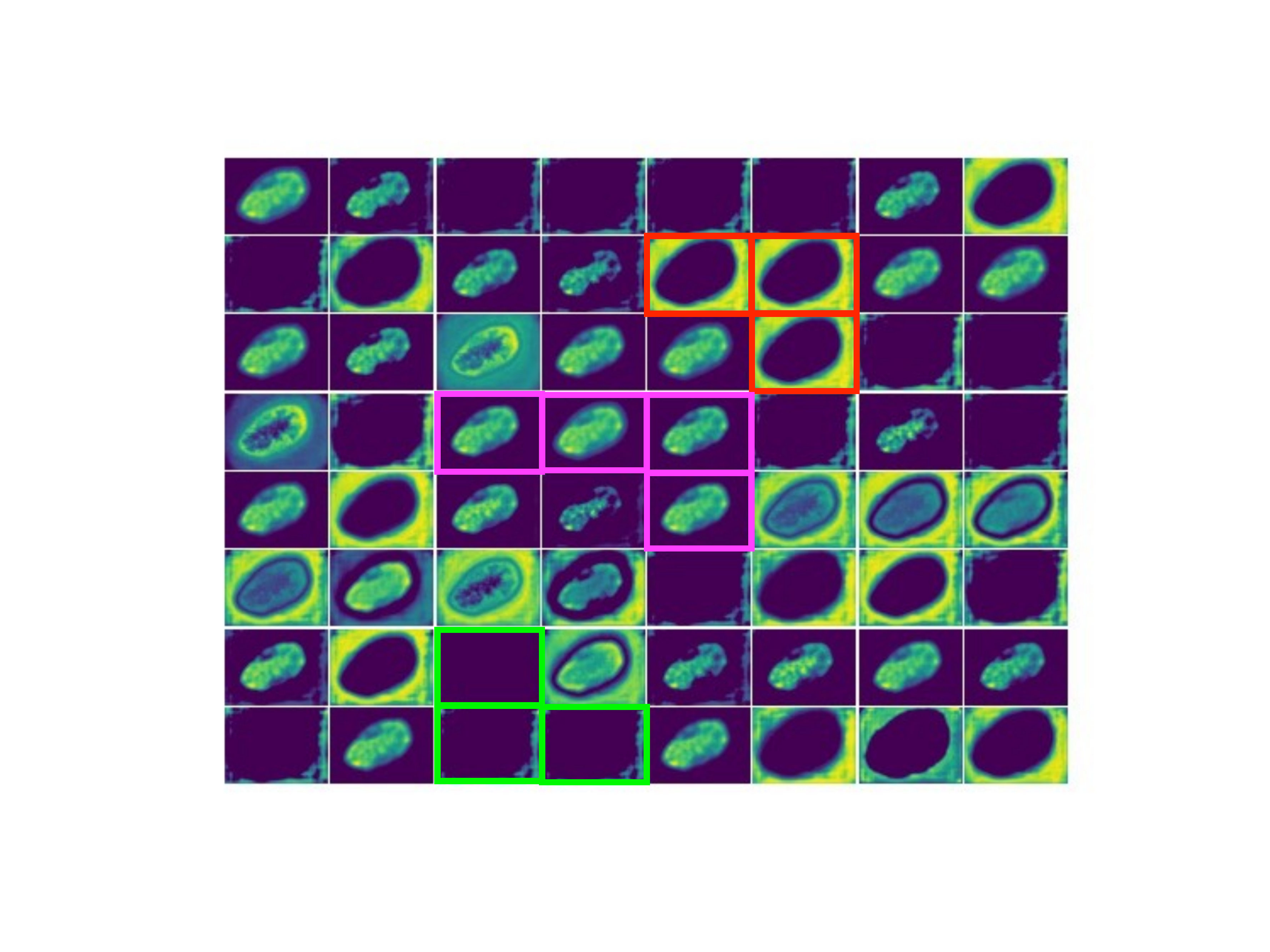}}
%  \vspace{1.5cm}
  \centerline{(c) Last Convolution Feature}\medskip
\end{minipage}
\caption{Visualization of Feature Redundancy.}
\label{Redundancy}
\end{figure}

\section{Methodology}
\label{sec:methodology}
\subsection{GP-module}
\label{GPmodule}

\begin{figure}[htb]
\begin{minipage}[b]{0.3\linewidth}
  \centering
  \centerline{\includegraphics[width=1.0\textwidth]{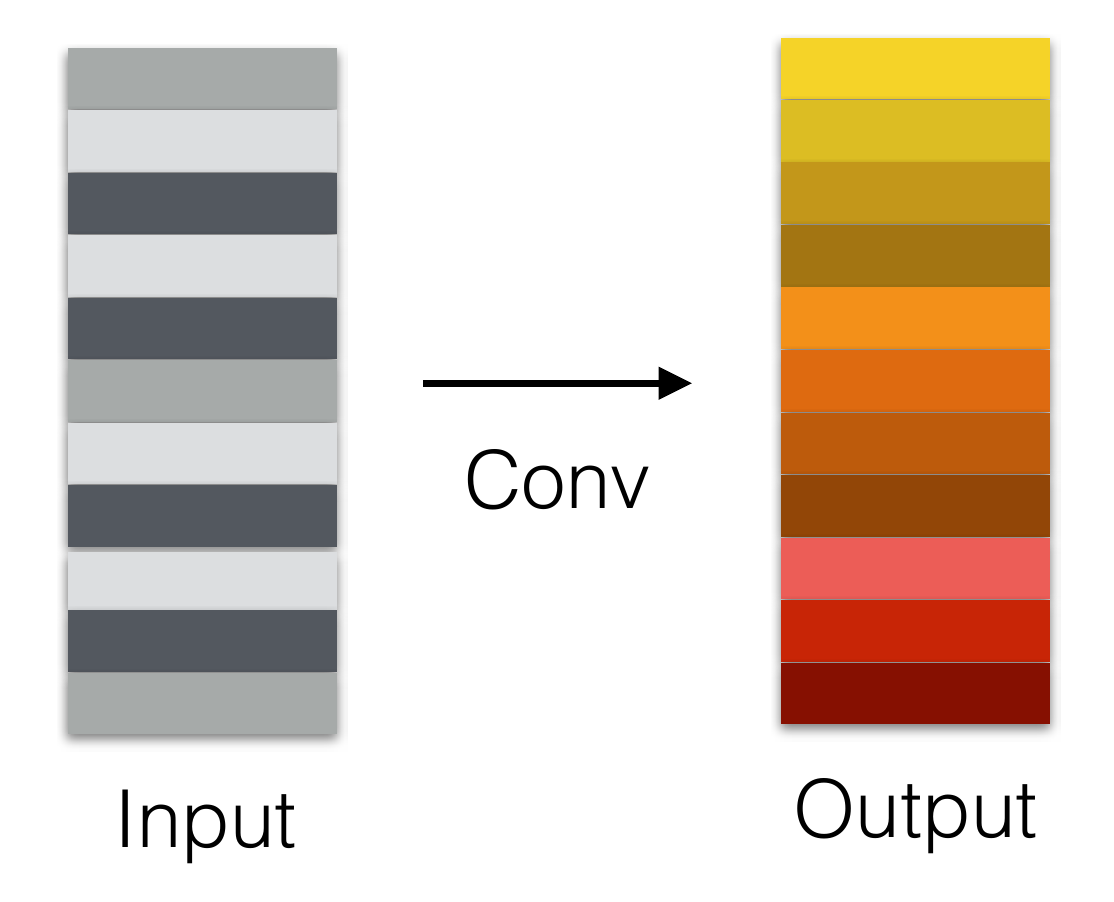}}
%  \vspace{2.0cm}
  \centerline{(a) Ordinary Convolution}\medskip
  %\label{modules1}
\end{minipage}
\begin{minipage}[b]{0.6\linewidth}
  \centering
  \centerline{\includegraphics[width=1.0\textwidth]{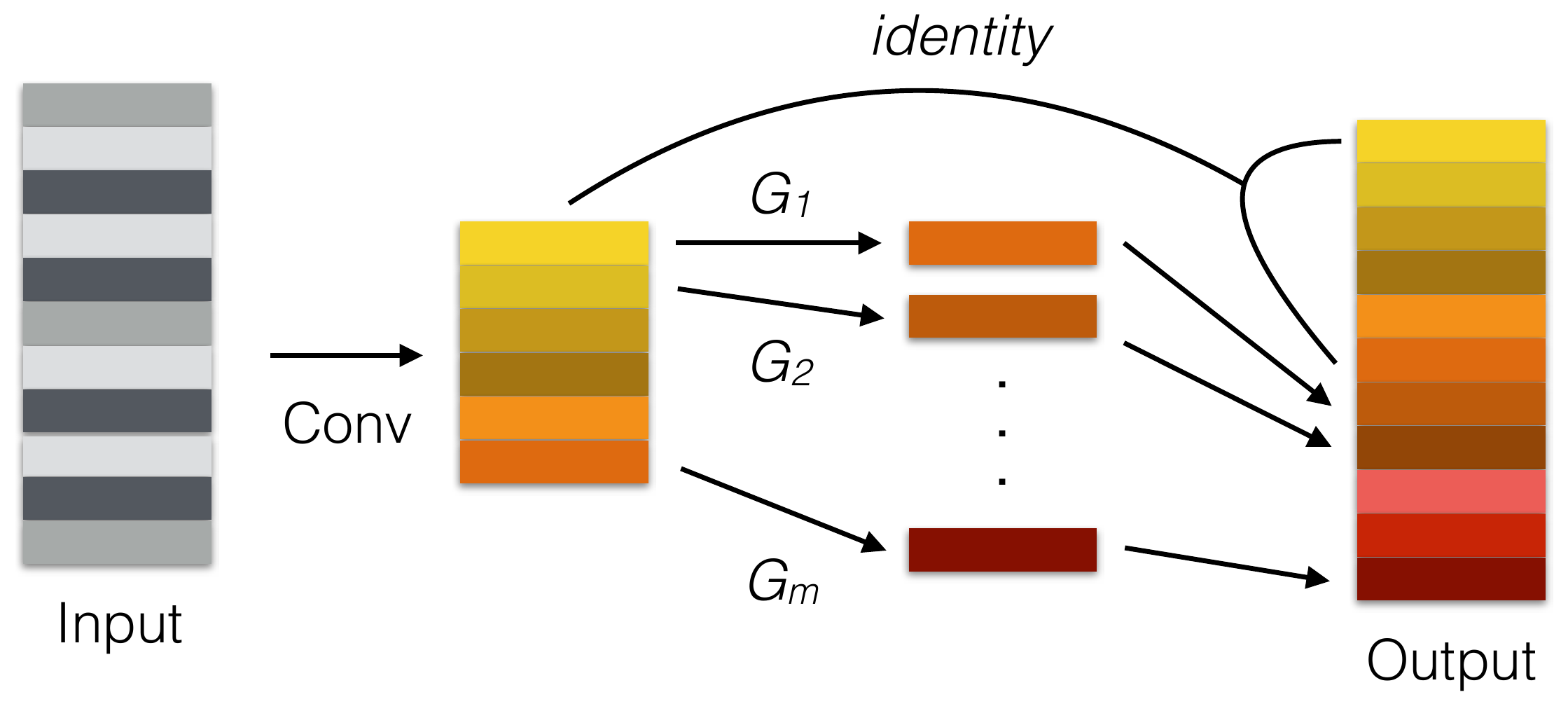}}
%  \vspace{1.5cm}
  \centerline{(b) Ghost Module}\medskip
  %\label{modules2}
\end{minipage}
\hfill
\begin{minipage}[b]{1.0\linewidth}
  \centering
  \centerline{\includegraphics[width=0.7\textwidth]{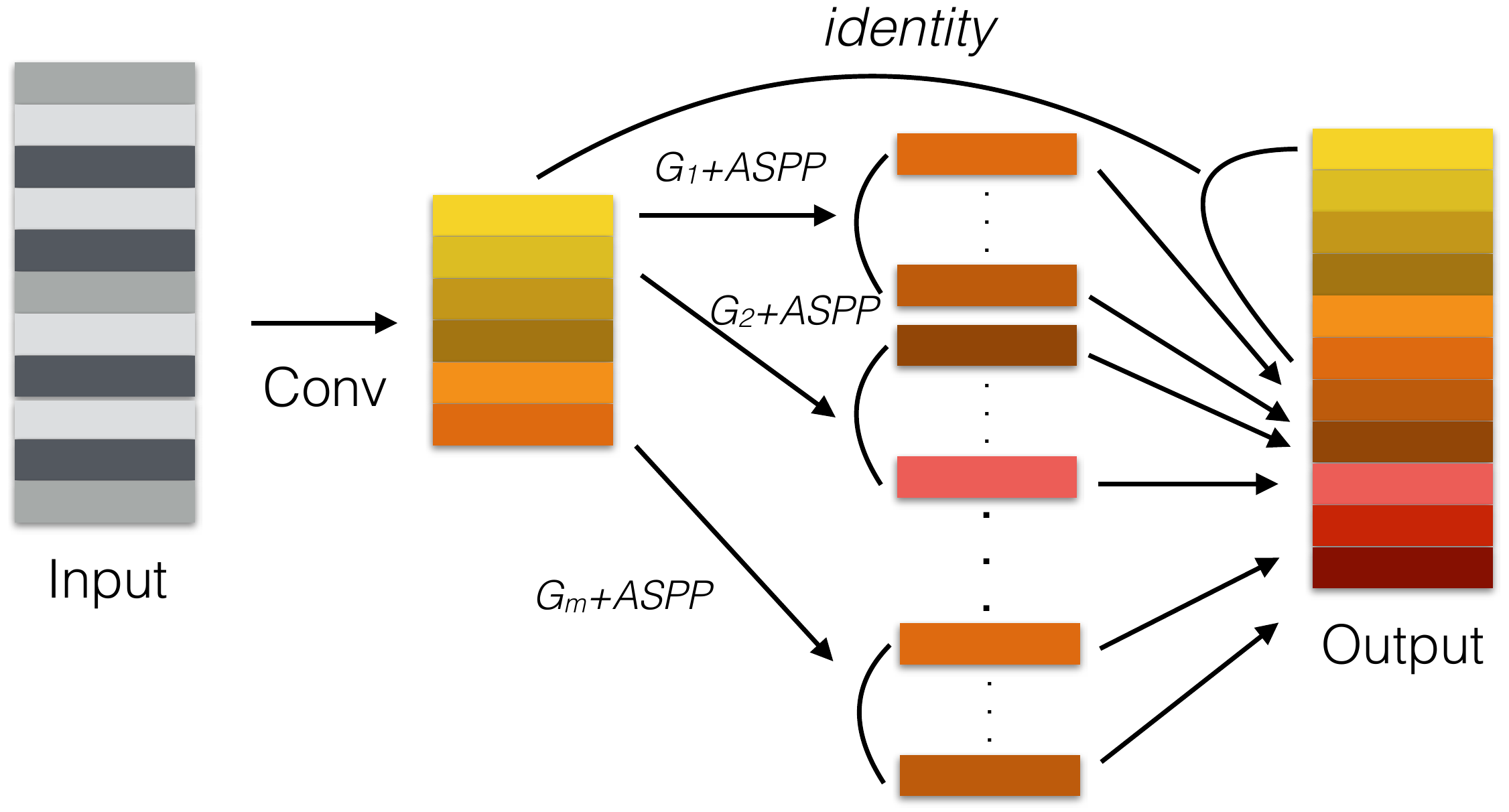}}
%  \vspace{1.5cm}
  \centerline{(c) GP-module}\medskip
  %\label{modules3}
\end{minipage}
\caption{Visualization of Feature Redundancy.}
\label{modules}
\end{figure}

U-Net and its existing variants have feature redundancy problem as shown in Fig~\ref{Redundancy}, where we visualize the first and last convolution layer of U-Net segmentation results. This redundancy includes two aspects. On the one hand, some feature maps contain little  or even no useful (green parts) information. On the other hand, there exist similar feature maps (as red, pink and cyan parts). Redundancy in feature maps can benefit model performance to some extent. However, the existing methods obtain sufficient feature maps at the expense of efficiency. To solve the problems, we propose GP-module, an enhanced version of ghost module that is first introduced in GhostNet~\cite{han2020ghostnet}. Ghost module uses cheap linear operations base on a handful of intrinsic feature maps to generate comparable features with traditional convolution operation as shown in Fig~\ref{modules}. To be specific, ordinary convolution operation can be formulated as $Y = X \ast f$, where $X\in\mathbf{R}^{c\times w \times h}$ is the input feature map with $c$ channels and $h$ and $w$ are the height and width, $\ast$ is the convolution operation, $f\in\mathbf{R}^{c\times k \times k \times n}$ is the convolution filters to produce $n$ feature maps and $k \times k$ is the kernel size, $Y\in\mathbf{R}^{w' \times h' \times n}$ is the output feature map with $n$ channels and $h'$ and $w'$ are the height and width. In a convolution operation, the number of parameters and FLOPs required can be calculated as Eq.\ref{eq2} and Eq.\ref{eq3}, respectively. Both of them can be very large when $c$ and $n$ are very large, which is the usual case.

\begin{equation}
N^{Para}_{conv} = c\cdot k \cdot k \cdot n\label{eq2}
\end{equation}

\begin{equation}
N^{FLOPs}_{conv} = c\cdot k \cdot k \cdot w' \cdot h' \cdot n\label{eq3}
\end{equation}

We can find out from Fig~\ref{Redundancy} that there are many similar features and they can be generate using fewer parameters and FLOPs. Ghost module~\cite{han2020ghostnet} deal with this problem by dividing feature maps into two parts. One part is a small number of intrinsic feature maps and the other part of feature maps is "ghosts" of the intrinsic feature maps. They are called ghost feature maps and are produced by using some cheap transformations based on the intrinsic feature maps. The idea is generating $m$ intrinsic feature maps and for each intrinsic feature map, applying several cheap linear operations to get $s$ ghost feature maps. By this way we obtain the $n = m \cdot s$ feature maps desired. The way to generate intrinsic feature maps is the same as ordinary convolution operation and the hyper-parameters (kernel size, padding, stride, etc.) are consistent with ordinary convolution to keep the same output spatial size ($Y' = X \ast f'$). The only difference is that $f'\in\mathbf{R}^{c\times k \times k \times m}$ and $Y'\in\mathbf{R}^{w' \times h' \times m}$, where $m \le n$ so the number of parameters can be greatly reduced.
After getting intrinsic feature maps, ghost features can be produced by applying a series of cheap operations on each intrinsic feature as Eq.\ref{eq5}:

\begin{equation}
y_{ij} = G_{ij}(y'_{i}), \quad \forall i = 1, ..., m, j = 1, ..., s\label{eq5}
\end{equation}
where $y'_{i}$ is the $i$-th intrinsic feature map of $Y'$ and $G_{ij}$ is the $j$-th linear operation (e.g. $3 \times 3$ and $5 \times 5$ linear kernels) applied on $y'_{i}$ to generate the $j$-th ghost feature map $y_{ij}$. Each $y'_{i}$ can get $s$ ghost feature maps except that the last one is the identity mapping for preserving the intrinsic feature map for each $y'_{i}$. So we can obtain $n = m \cdot s$ output feature maps $Y = [y_{11},y_{12}, ..., y_{1s}, ...y_{ms}]$. Note that the input and output of $G_{ij}$ are all single-channel feature maps, Eq.\ref{eq5} can be easily implemented by using depth-wise convolution. Ghost module can produce sufficient ghost features by selecting $s$ and kernel size in $G_{ij}$. But it uses the same single size kernel in all the $s$ linear operations for each intrinsic feature map $y'_{i}$, which is not conducive to generating diverse and informative feature maps. In order to better handle this problem, we introduce atrous spatial pyramid pooling (ASPP) mechanism  proposed in ~\cite{chen2017deeplab} and get our GP-module as shown in Fig~\ref{modules}. Our GP-module can be expressed as Eq.\ref{eq6}:

\begin{equation}
y_{ij} = G'_{i}(y'_{i}, j), \quad \forall i = 1, ..., m, j = 1, ..., s\label{eq6}
\end{equation}

$G'_{i}$ is explicitly related to $j$, which means for each intrinsic feature map $y'_{i}$, when generating more than $s > 2$ ghost feature maps, GP-module applies the same kernel size but different dilation rates (except the identity mapping). By introducing ASPP, GP-module can expand the receptive field and capture multi-scale contextual information so as to generate heterogeneous representative ghost feature maps without additional parameters and FLOPs given that GP-module uses the same kernel size as ghost module. For each ghost module/GP-module, there is one ordinary convolution operation to generate $m$ intrinsic feature maps and $m \cdot (s-1) = \frac{n}{s} \cdot (s-1)$ linear operations (one of the $s$ is the identity mapping operation) to get ghost feature maps. Note the kernel size in linear operations is $d \times d$ and the number of parameters and FLOPs required in ghost module/GP-module can be calculated as Eq.\ref{eq7} and Eq.\ref{eq8}, respectively. In practice, $d \times d$ and $k \times k$ have the similar magnitude and $s$ can be much smaller than $c$. So the parameters compression ratio and FLOPs acceleration ratio can be calculated as Eq.\ref{eq9} and Eq.\ref{eq10}, respectively.

\begin{equation}
N^{Para}_{gp} = c \cdot k \cdot k \cdot \frac{n}{s} + \frac{n}{s} \cdot (s-1) \cdot d \cdot d \label{eq7}
\end{equation}

\begin{equation}
N^{FLOPs}_{gp} =  c \cdot k \cdot k \cdot \frac{n}{s} \cdot w' \cdot h' + \frac{n}{s} \cdot (s-1) \cdot d \cdot d \cdot w' \cdot h' \label{eq8}
\end{equation}

\begin{equation}
r_{P} =  \frac{N^{Para}_{conv}}{N^{Para}_{gp}} \approx \frac{c \cdot s \cdot k \cdot k}{c \cdot k \cdot k + (s-1) \cdot d \cdot d} \approx \frac{c \cdot s}{c+s-1} \approx s \label{eq9}
\end{equation}

\begin{equation}
r_{F} =  \frac{N^{FLOPs}_{conv}}{N^{FLOPs}_{gp}} \approx \frac{c \cdot s}{c+s-1} \approx s \label{eq10}
\end{equation}
The parameters compression ratio is equal to FLOPs acceleration ratio which means our GP-module does not cost extra computing resources based on ghost module and greatly reduced parameters and FLOPs compared to ordinary convolution operation.

% \begin{figure}[htbp]
% \centering
% \subfigure[Ordinary Convolution]{
% \includegraphics[width=0.3\textwidth]{imgs/fig1.pdf}
% \label{modules1}
% }
% \subfigure[Ghost Module]{
% \includegraphics[width=0.65\textwidth]{imgs/fig2.pdf}
% \label{modules2}
% }

% \subfigure[GP-module]{
% \includegraphics[width=0.75\textwidth]{imgs/fig3.pdf}
% \label{modules3}
% }
% \caption{Diagram of the three modules}
% \label{modules}
% %\vspace{-0.5cm}
% \end{figure}

\subsection{Network Architecture}
\begin{figure}[htbp]
\begin{minipage}[b]{0.1\linewidth}
  \centering
  \centerline{\includegraphics[width=1.0\textwidth]{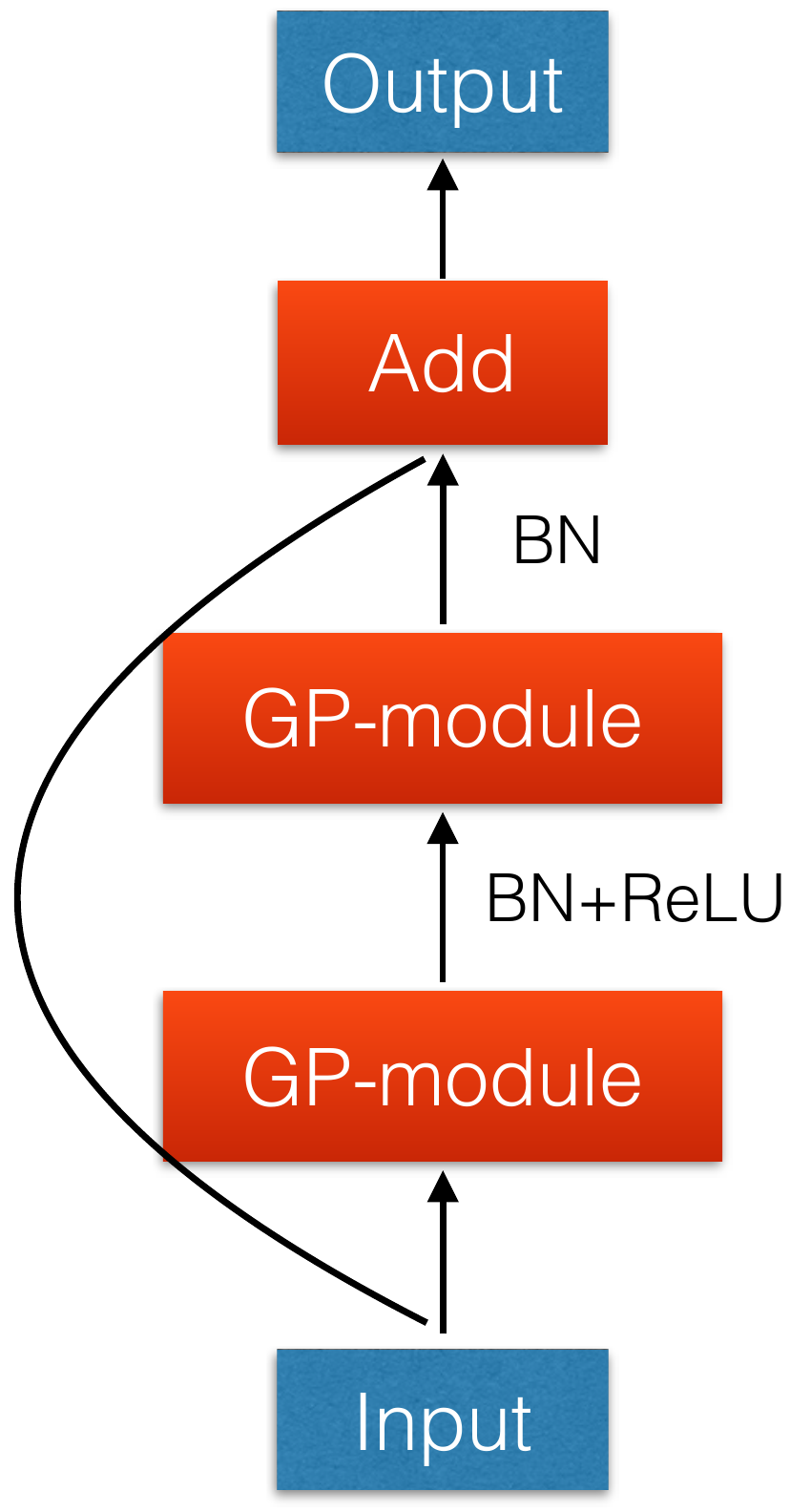}}
%  \vspace{1.5cm}
  \centerline{(a) GP-bneck}\medskip
\end{minipage}
\hfill
\begin{minipage}[b]{0.9\linewidth}
  \centering
  \centerline{\includegraphics[width=0.7\textwidth]{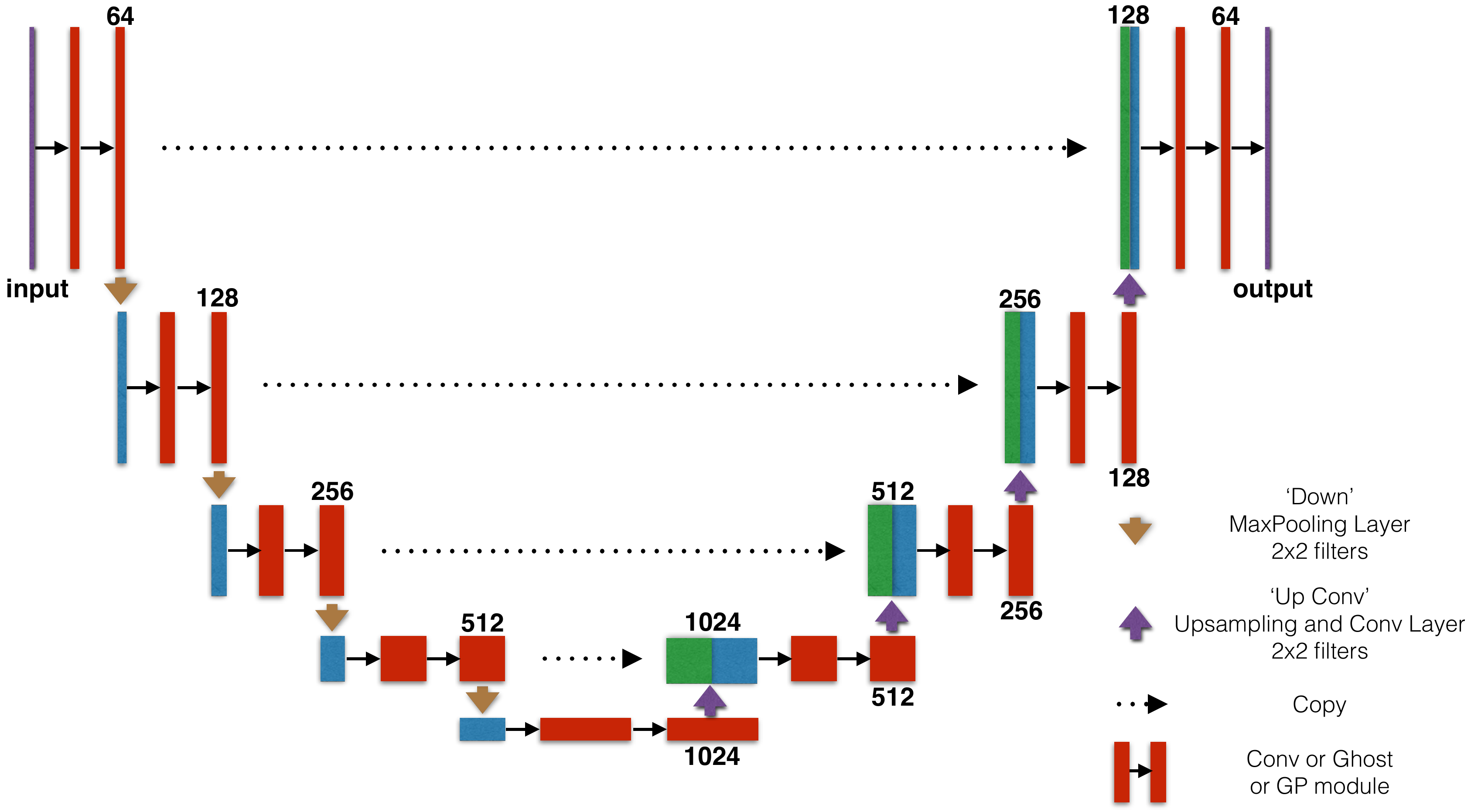}}
%  \vspace{1.5cm}
  \centerline{(b) U-Net \& Ghost U-Net \& GPU-Net}\medskip
\end{minipage}
\caption{Visualization of Feature Redundancy.}
\label{net}
\end{figure}

Based on GP-module, we introduce GP-bottleneck (GP-bneck) as shown in Fig~\ref{net}, which is similar as G-bneck in~\cite{han2020ghostnet}. GP-bottleneck has a residual structure proposed in ResNet~\cite{he2016deep} and two stacked GP-modules, which corresponds to the two continuous convolution operations of each level in U-Net. Batch normalization (BN)~\cite{ioffe2015batch} is used after each GP-module and ReLU is used only after the first GP-module in GP-bottleneck. After getting GP-bottleneck, we replace the convolution operations in U-Net with our GP-bottleneck and name it GPU-Net, which is a lightweight and powerful (like its name) network for medical image segmentation as shown in Fig~\ref{net}, where the red module indicates the location of the replacement. We use binary cross entropy as loss function to train the whole network.

\section{Experiments and Results}
\label{sec:experiments}
\subsection{Dataset}
\label{sec:dataset}

\textbf{Skin Lesion Segmentation}
Skin cancer is the most common cancer and accurate predictions of lesion segmentation boundaries in dermoscopic images can benefit clinical treatment. The dataset is from MICCAI 2018 Workshop~\cite{codella2019skin} (ISIC for short). ISIC dataset contains 2594 images in total and is split into training set (1814 samples, $\sim 70\%$ ), validation set (260 samples, $\sim 10\%$), and test set (520 samples, $\sim 20\%$). All the samples are resized into $192 \times 256$ given that the original samples are slightly different.

\textbf{Lung Segmentation}
Lung segmentation is important for analyzing lung
related diseases, and can be further applied into lung lesion segmentation and other problems. This dataset is from Kaggle Challenge (LUNA for short). LUNA dataset contains 267 2D CT images and is split into training set (186 samples, $\sim 70\%$ ), validation set (27 samples, $\sim 10\%$), and test set (54 samples, $\sim 20\%$). The original image size is $512 \times 512$ and we resize all the images into $256 \times 256$ in the experiment.

\textbf{Nuclei Segmentation}
Identifying the cells’ nuclei is the starting point of many research and can help researchers better understand the underlying biological processes. This dataset comes from Kaggle 2018 Data Science Bowl (DSB for short). DSB dataset has 670 nucleus images and corresponding masks. The whole dataset is split into training set (469 samples, $70\%$ ), validation set (67 samples, $10\%$), and test set (134 samples, $20\%$). The original image size is $96 \times 96$ and remain the same in the experiment.

\begin{equation}
AC=\frac{TP+TN}{TP+TN+FP+FN}\label{con:ac}
\end{equation}

%\end{small}
%\begin{small}
\begin{equation}
F1=2 \frac{\left|GT\cap{SR}\right|}{\left|GT\right|+\left|SR\right|},
JS=\frac{\left|GT\cap{SR}\right|}{\left|GT\cup{SR}\right|}\label{con:js} 
\end{equation}
%\end{small}
%\begin{small}
% \begin{equation}
% JS=\frac{\left|GT\cap{SR}\right|}{\left|GT\cup{SR}\right|}\label{con:js}
% \end{equation}
%\end{small}

\begin{table*}[!h]
    \caption{Experimental Results on Three Datasets}
    %\vspace{20pt}
    \centering
    %\begin{tabular}{p{0.7cm}p{0.7cm}p{0.7cm}p{0.7cm}p{0.7cm}p{0.7cm}p{0.7cm}p{0.7cm}}
    \begin{tabular}{ccccccc}
        \hline
        Dateset& Methods& AC& F1& JS& Params(M)& FLOPs(G)\\
        \hline
        ISIC  & U-Net & 0.9554 & 0.8791 & 0.8071 & 34.53 & 49.10\\
        & Ghost U-Net & 0.9605 & 0.8852 & 0.8140 & 9.31 & 18.78\\
        & GPU-Net & \textbf{0.9613} & \textbf{0.8926} & \textbf{0.8232} & 8.27 & 17.56\\
        \hline
        LUNA  & U-Net & 0.9813 & 0.9785 & 0.9644 & 34.53 & 65.47\\
        & Ghost U-Net & 0.9879 & 0.9796 & 0.9665 & 9.31 & 25.05\\
        & GPU-Net & \textbf{0.9892} & \textbf{0.9811} & \textbf{0.9675} & 8.27 & 23.42\\
        \hline
        DSB  & U-Net & 0.9719 & 0.8698 & 0.7944 & 34.53 & 9.21\\
        & Ghost U-Net & 0.9716 & 0.8731 & 0.7954 & 9.31 & 3.52\\
        & GPU-Net & \textbf{0.9727} & \textbf{0.8819} & \textbf{0.8049} & 8.27 & 3.29\\
        \hline       
    \end{tabular}
    \label{table1}
\end{table*}

\subsection{Experimental Setup}
We implement all the experiments on a NVIDIA TITAN V GPU. We use a batch size of 4 for ISIC dataset, 2 for LUNA dataset and 16 for DSB dataset since they have different input sizes. We train all the models for 100 epochs and the initial learning rate is $0.001$. We compare our GPU-Net with original U-Net and Ghost U-Net (replace convolution operations in U-Net with G-bneck in~\cite{han2020ghostnet}). We set $s=2$ and $d=3$ in ghost module of G-bneck, which can achieve best performance according to~\cite{han2020ghostnet}. For our GP-module, we set $s=6$, which means $s-1=5$ cheap operations. We set the same $d=3$ for 4 of the 5 cheap operations and their dilation rates are 1, 6, 12, 18, respectively. The last cheap operation is $1 \times 1$ depth-wise convolution. The other hyper-parameters in GPU-Net/Ghost U-Net are consistent with U-Net. Our code is available at https://github.com/Heng14/GPU-Net.

\begin{figure}[htbp]
\begin{minipage}[b]{1.0\linewidth}
  \centering
  \centerline{
    \includegraphics[width=0.2\textwidth]{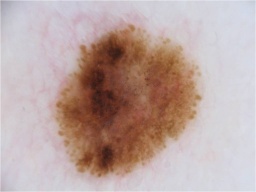}
    \includegraphics[width=0.2\textwidth]{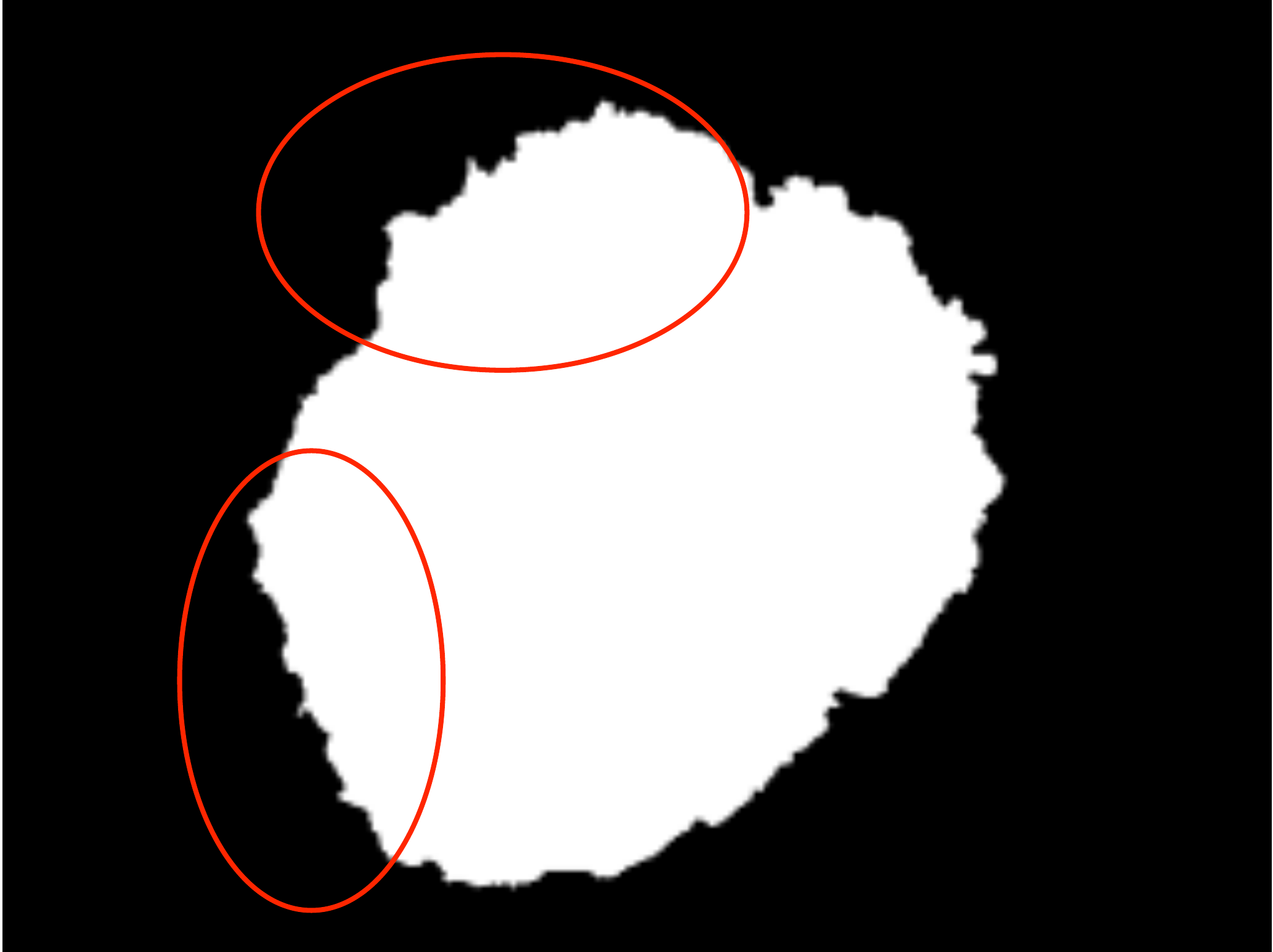}
    \includegraphics[width=0.2\textwidth]{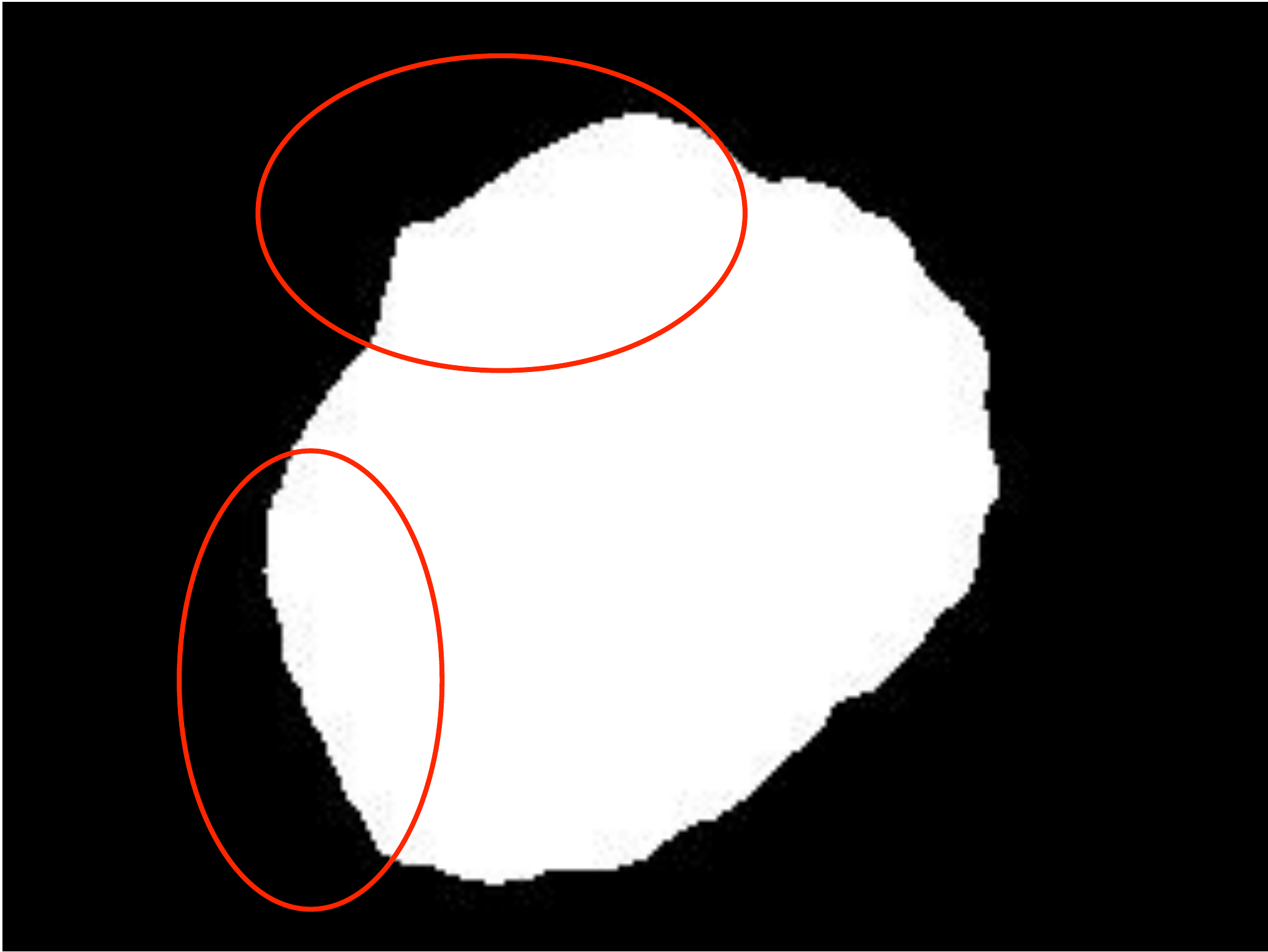}
    \includegraphics[width=0.2\textwidth]{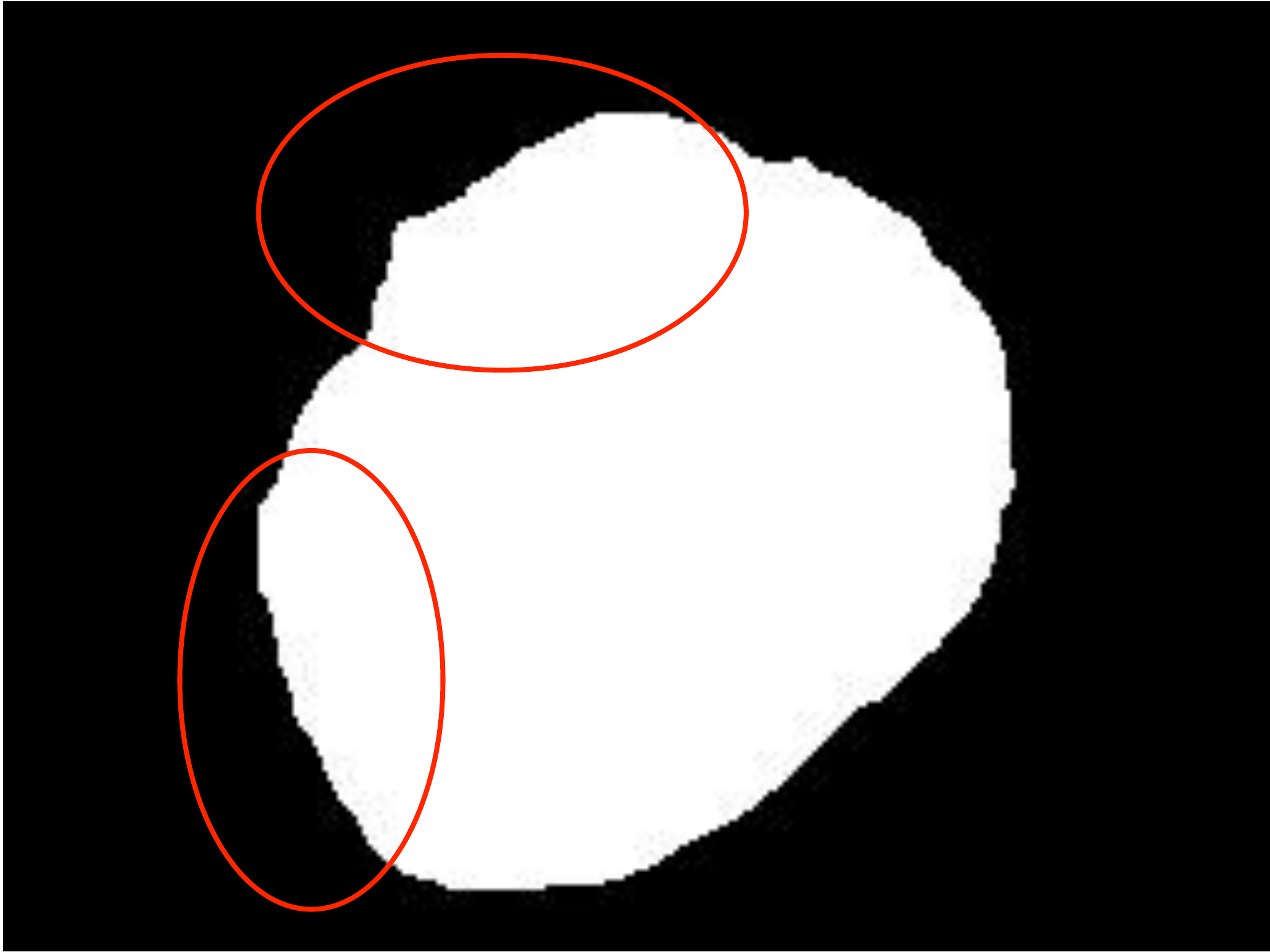}
    \includegraphics[width=0.2\textwidth]{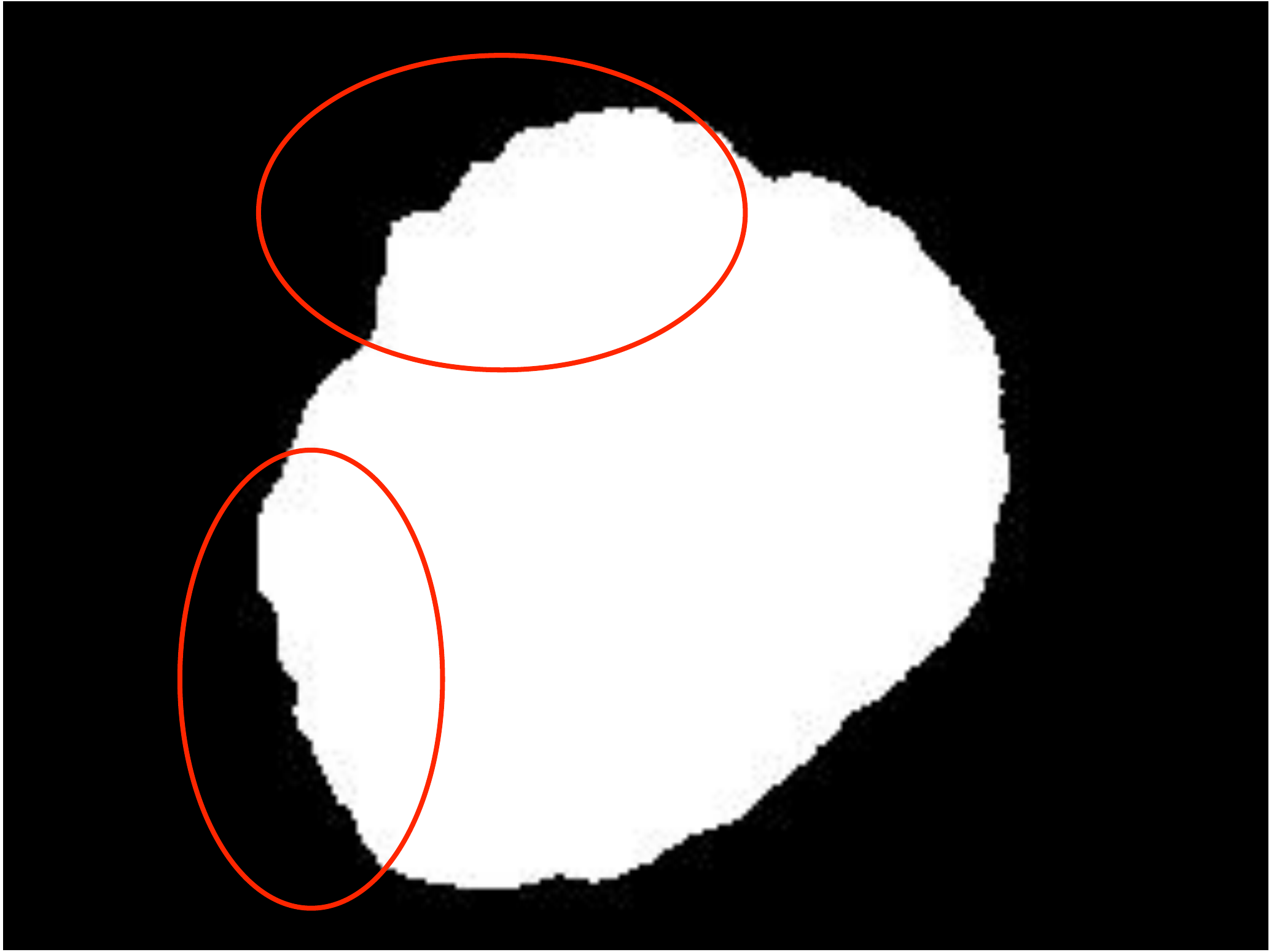}
  }
%  \vspace{1.5cm}
  \centerline{(a) ISIC}\medskip
\end{minipage}
%\hfill
\begin{minipage}[b]{1.0\linewidth}
  \centering
  \centerline{
    \includegraphics[width=0.2\textwidth]{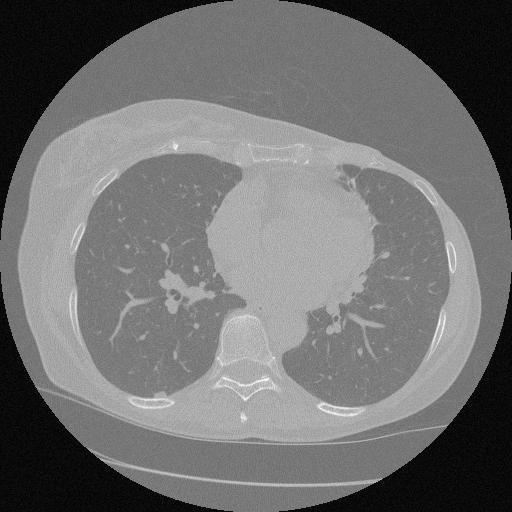}
    \includegraphics[width=0.2\textwidth]{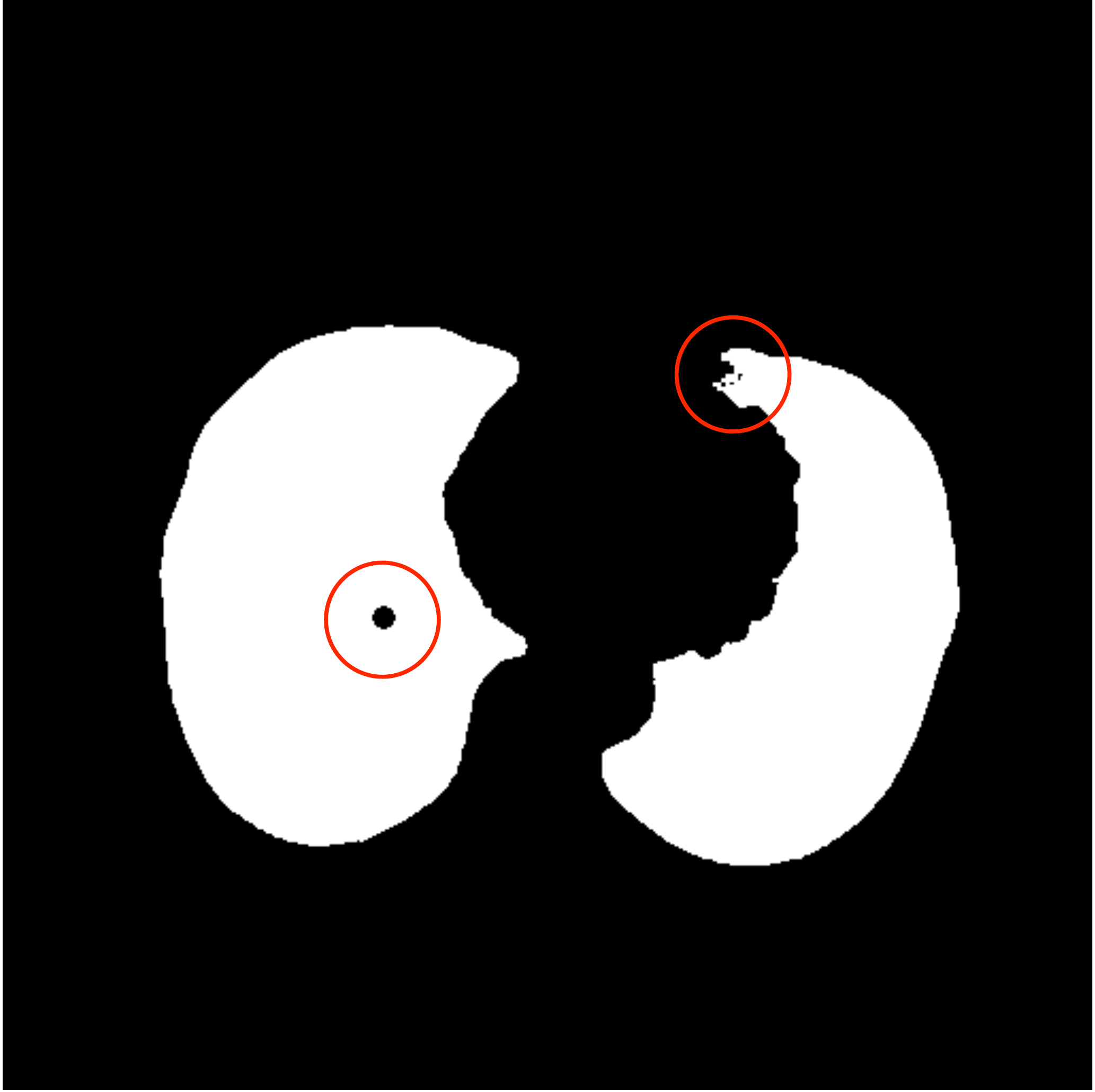}
    \includegraphics[width=0.2\textwidth]{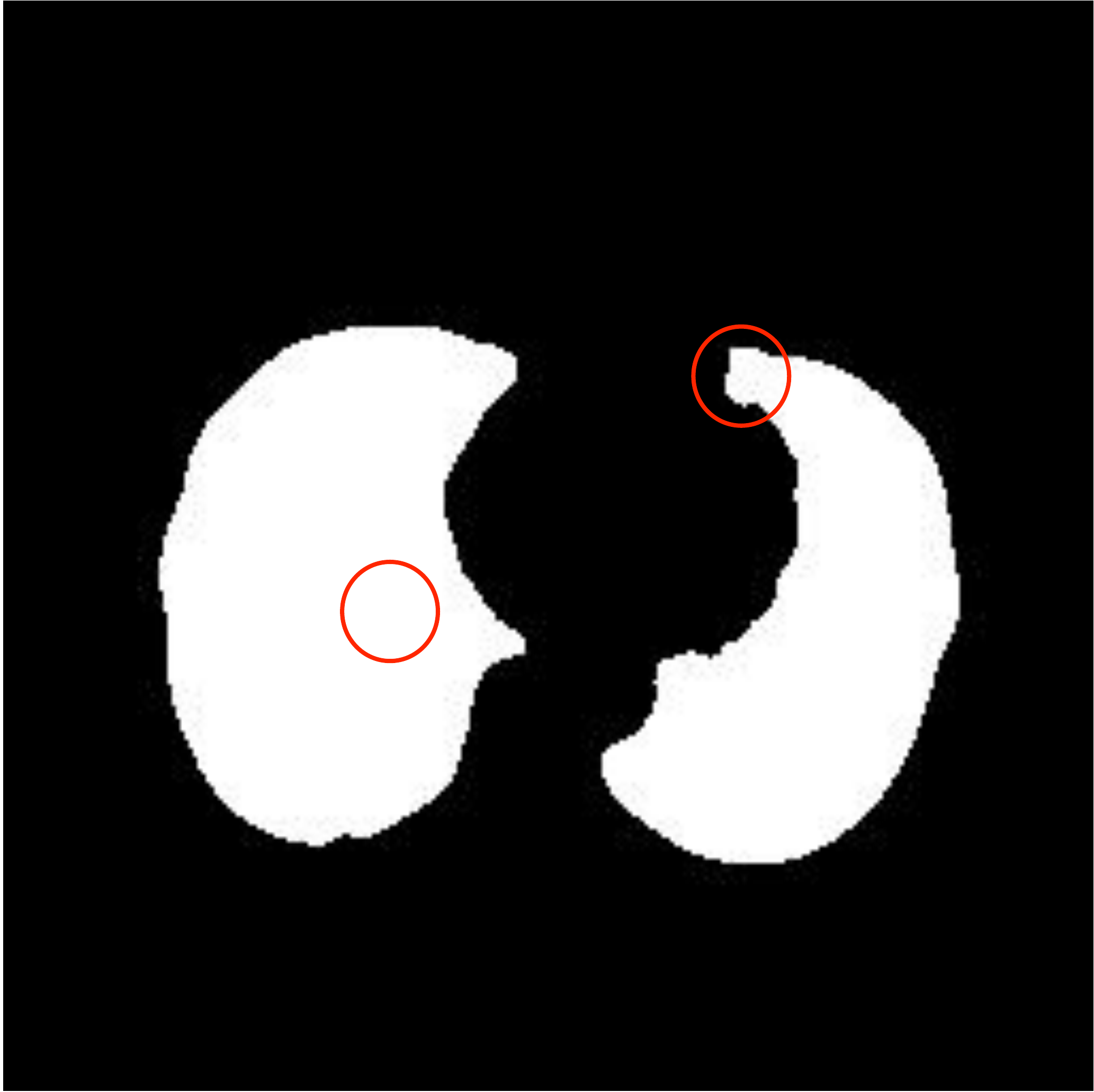}
    \includegraphics[width=0.2\textwidth]{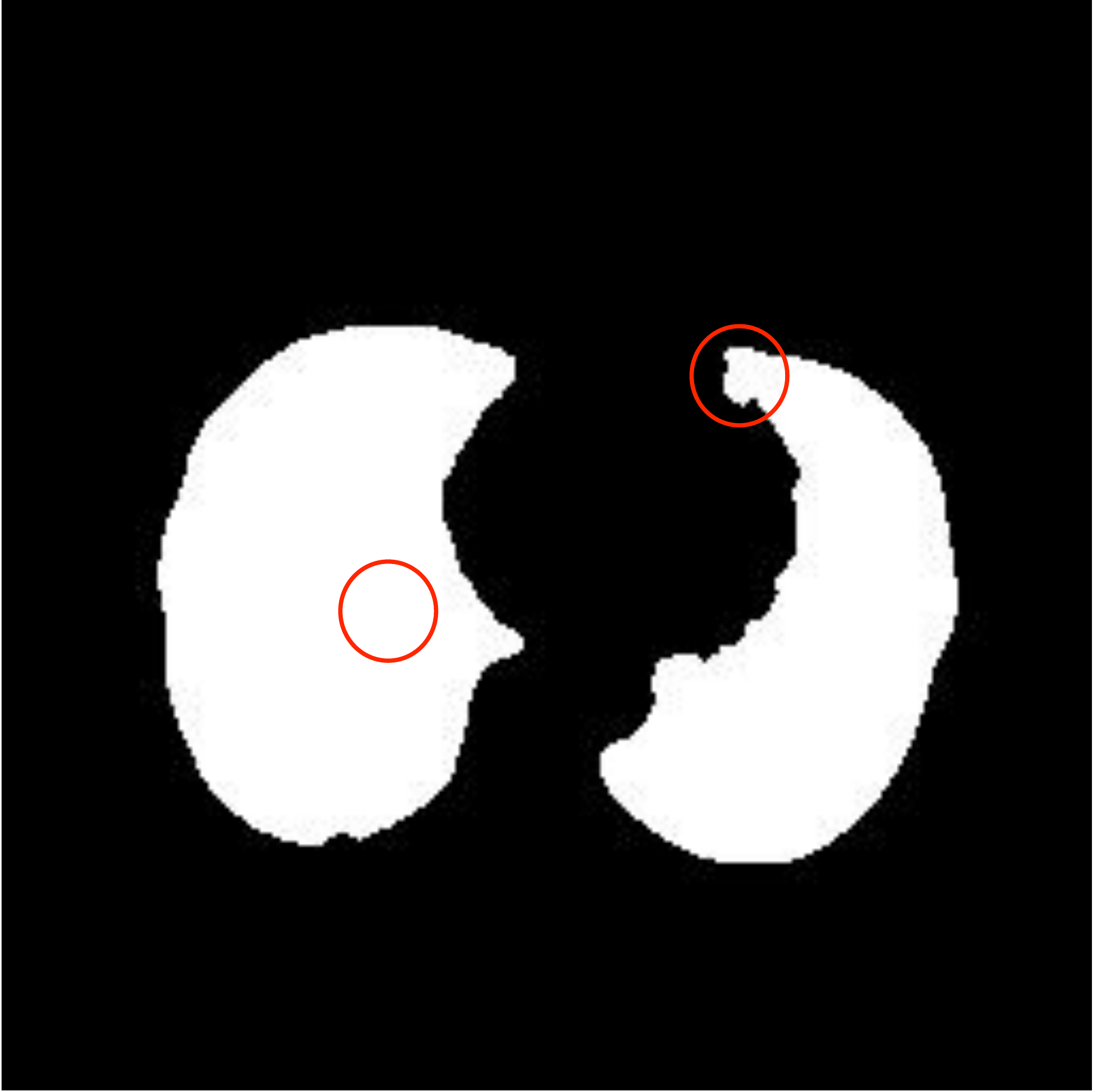}
    \includegraphics[width=0.2\textwidth]{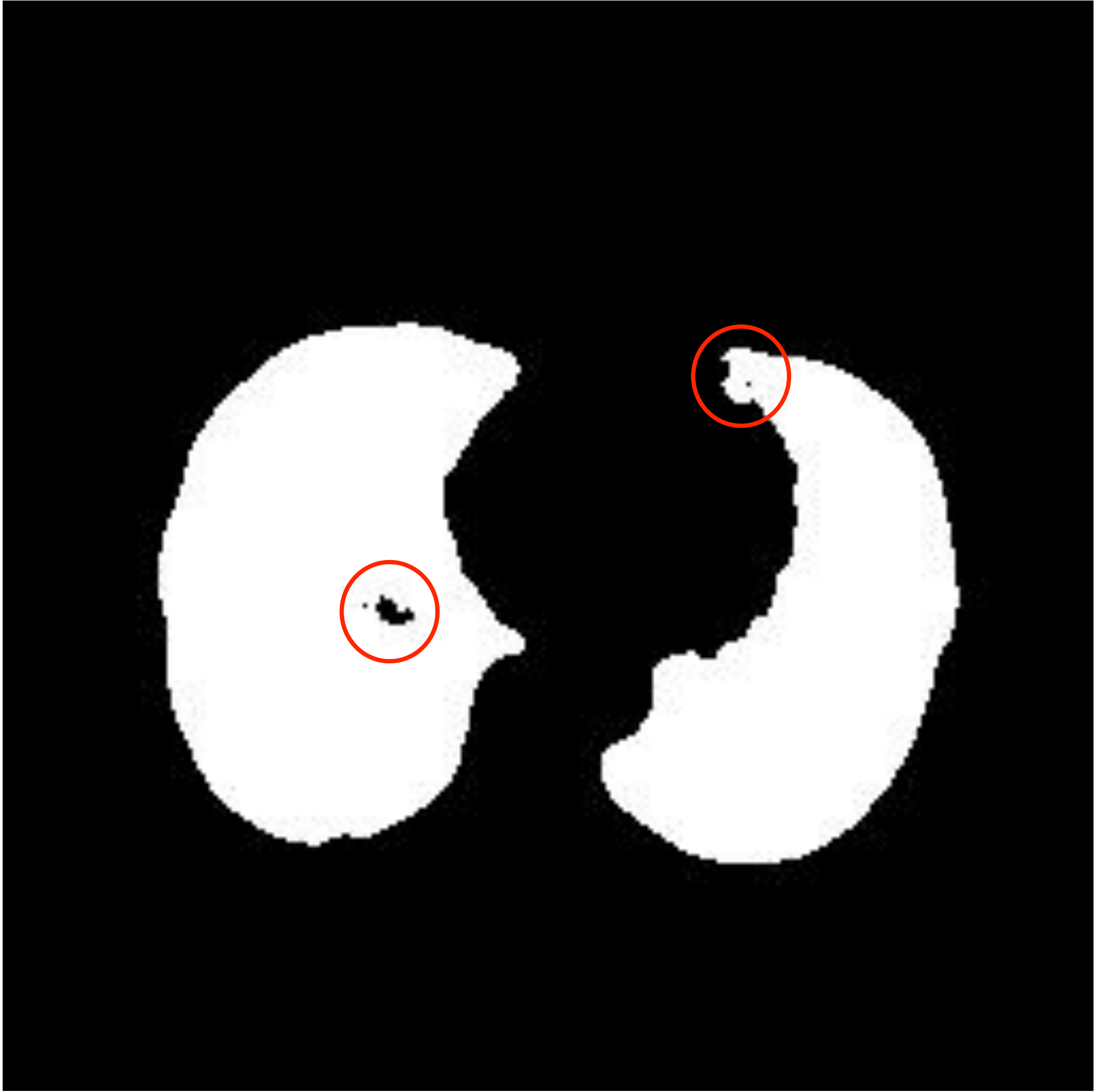}
  }
%  \vspace{1.5cm}
  \centerline{(b) LUNA}\medskip
\end{minipage}
\begin{minipage}[b]{1.0\linewidth}
  \centering
  \centerline{
    \includegraphics[width=0.2\textwidth]{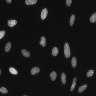}
    \includegraphics[width=0.2\textwidth]{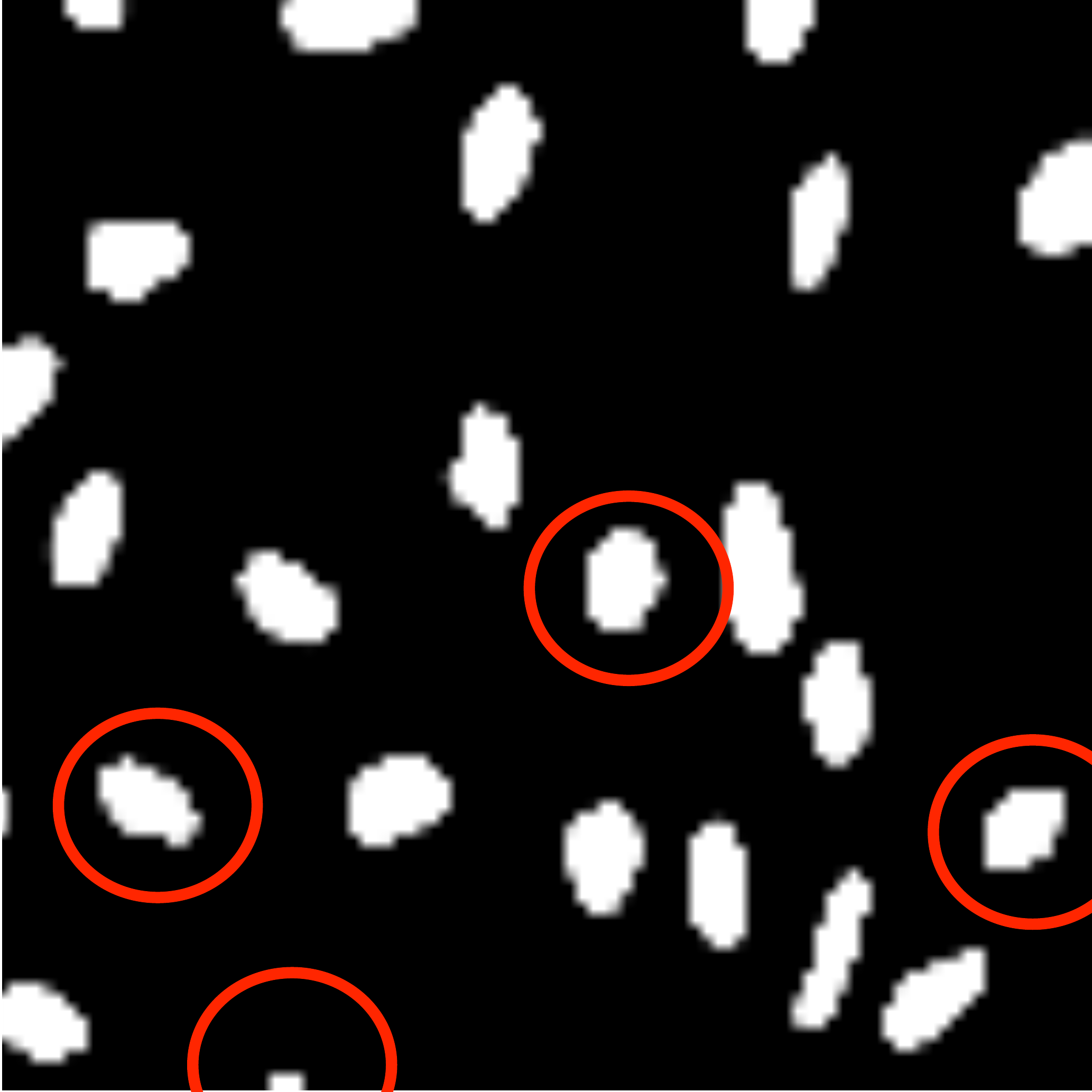}
    \includegraphics[width=0.2\textwidth]{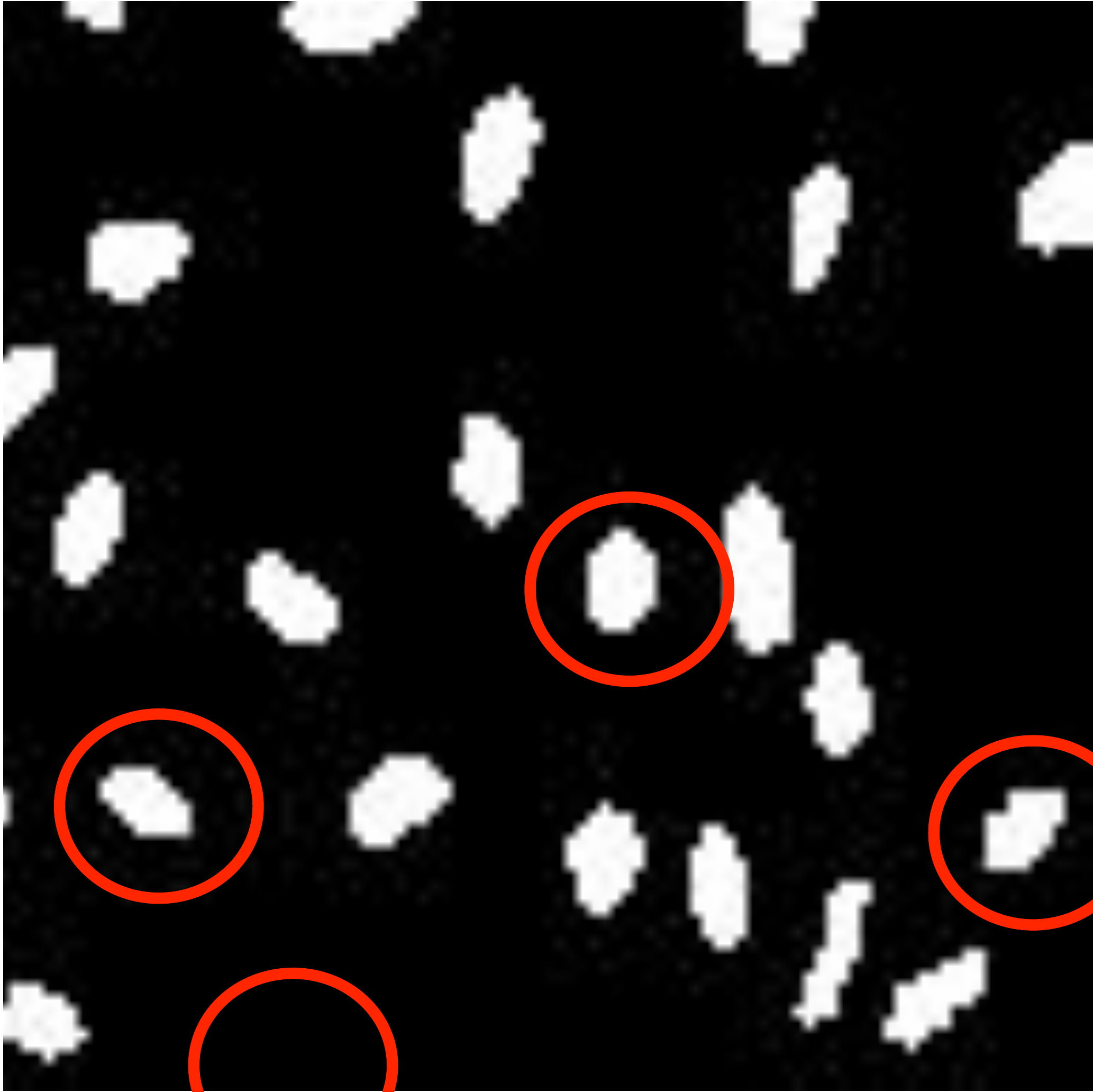}
    \includegraphics[width=0.2\textwidth]{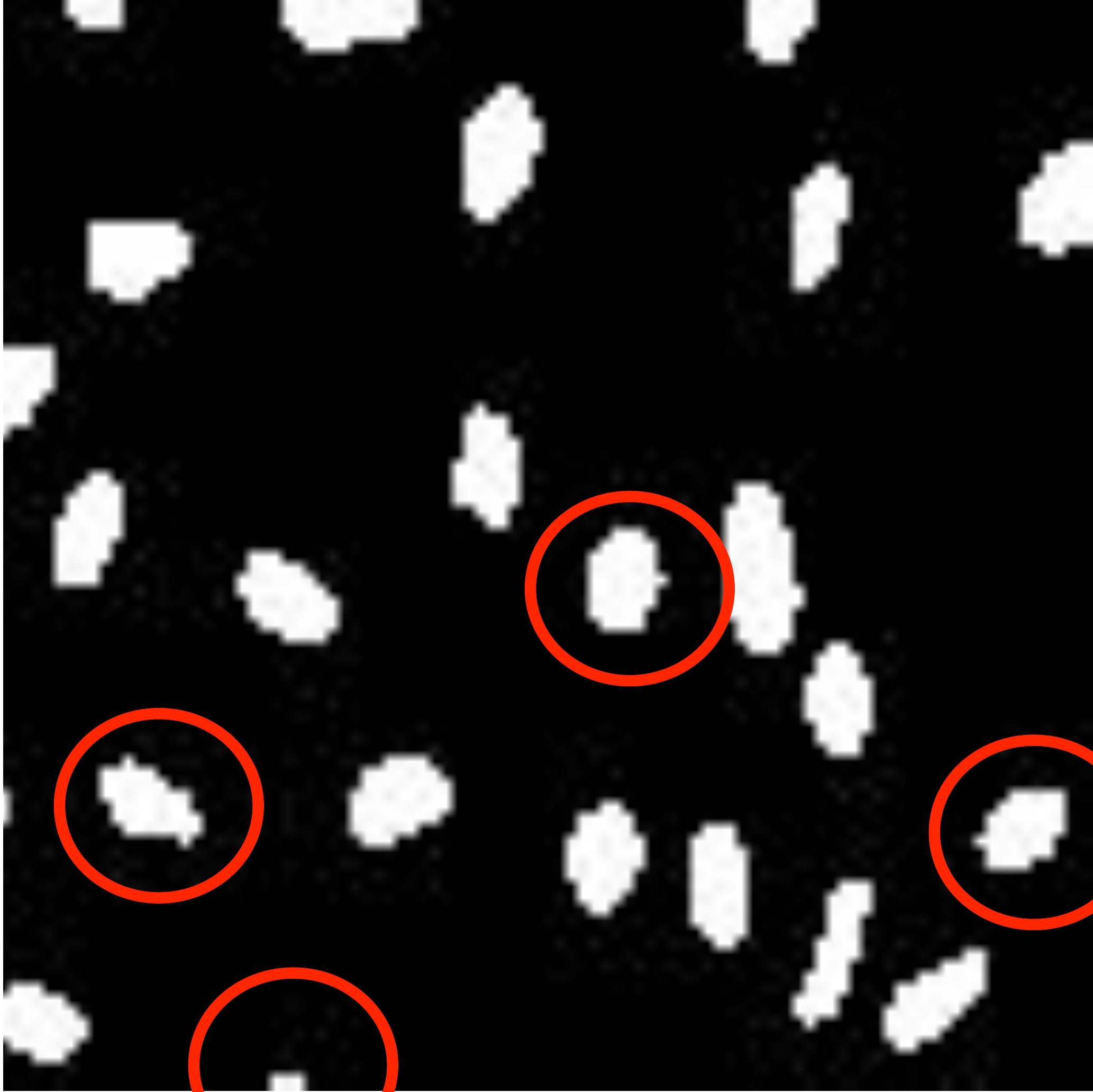}
    \includegraphics[width=0.2\textwidth]{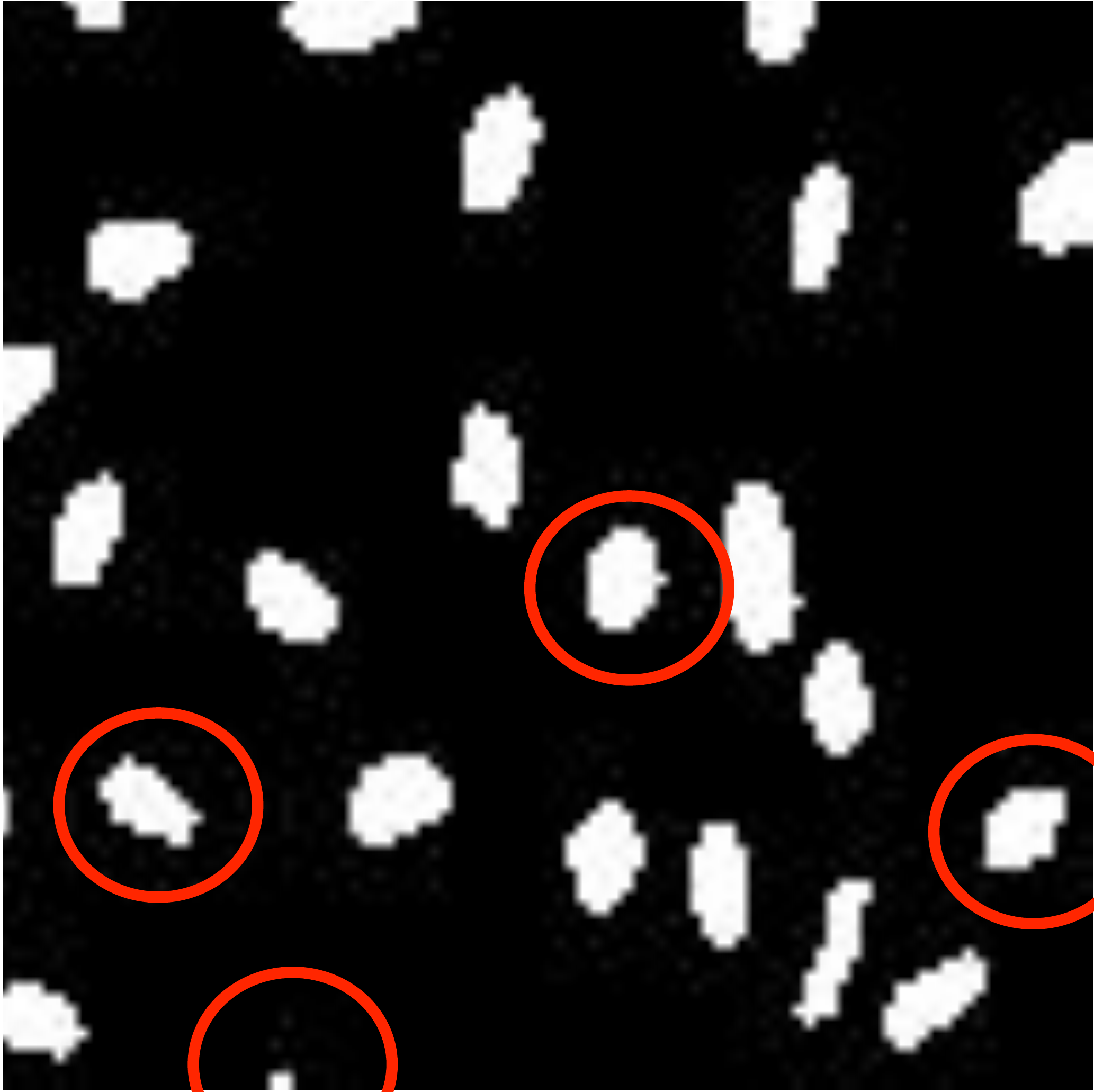}
  }
%  \vspace{1.5cm}
  \centerline{(c) DSB}\medskip
\end{minipage}
\vspace{-0.7cm}
\caption{Segmentation result samples. From left to right: input images, ground truth, U-Net, Ghost U-Net and GPU-Net outputs.}
\label{segres}
\end{figure}

\subsection{Results and Discussion}
To make a detailed comparison of the model performance, we take Accuracy (AC, Eq.\ref{con:ac}), F1-score and Jaccard similarity (F1 and JS, Eq.\ref{con:js}) as quantitative analysis metrics. Variables involved in these formulas are: True Positive (TP), True Negative (TN), False Positive (FP), False Negative (FN), Ground Truth(GT) and Segmentation Result (SR).

\begin{figure}[htbp]
\begin{minipage}[b]{1.0\linewidth}
  \centering
  \centerline{\includegraphics[width=0.45\textwidth]{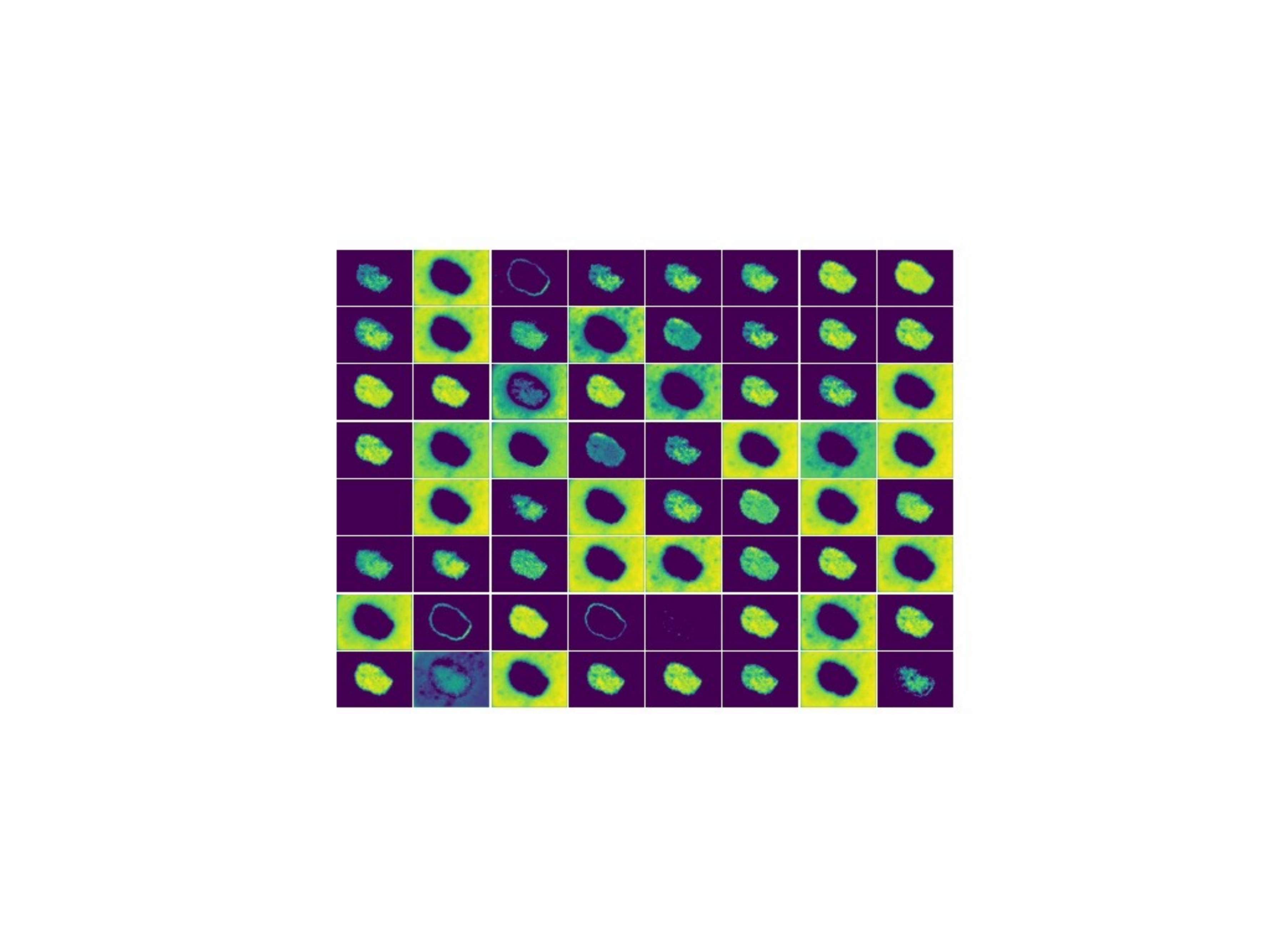}}
%  \vspace{2.0cm}
  \centerline{(a) U-Net}\medskip
\end{minipage}
\begin{minipage}[b]{0.45\linewidth}
  \centering
  \centerline{\includegraphics[width=1.0\textwidth]{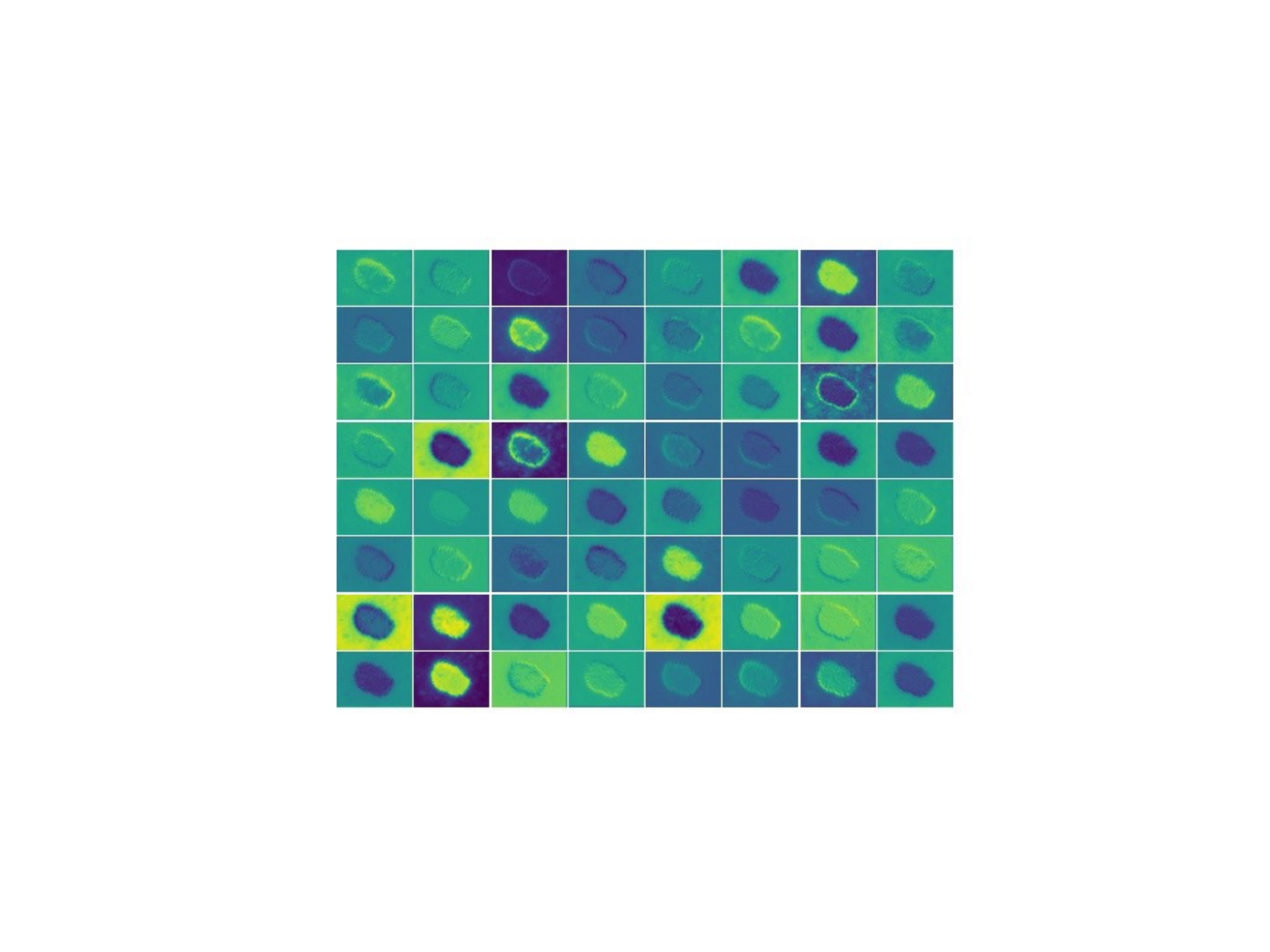}}
%  \vspace{1.5cm}
  \centerline{(b) Ghost U-Net}\medskip
\end{minipage}
\hfill
\begin{minipage}[b]{0.45\linewidth}
  \centering
  \centerline{\includegraphics[width=1.0\textwidth]{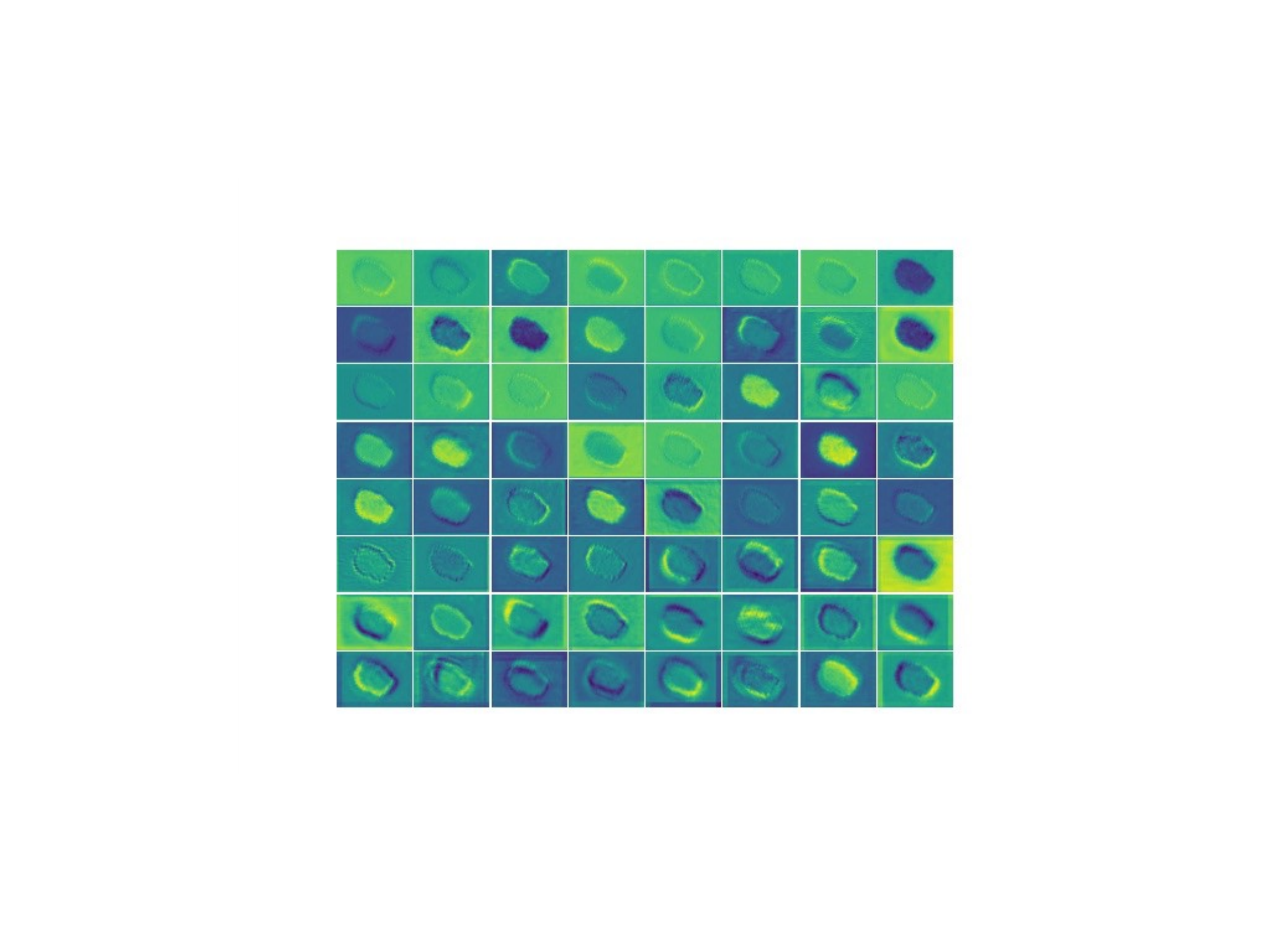}}
%  \vspace{1.5cm}
  \centerline{(c) GPU-Net}\medskip
\end{minipage}
\vspace{-0.3cm}
\caption{Visualization of Feature Maps.}
\label{feature}
\end{figure}

Experimental results are listed as Table~\ref{table1} and our method can achieve better segmentation performance with significantly reduced parameters and FLOPs. We also show some segmentation results in Fig~\ref{segres}, where we can find that our GPU-Net can better capture details and have better segmentation results at the edges (red circles). To better demonstrate the important role of our GP-module, we visualize the feature maps in the first level of the three networks as in Fig~\ref{feature}. We can see that U-Net has many similar and blank feature maps while Ghost U-Net and our GPU-Net can make full use of all feature maps. Our GPU-Net can learn more diverse feature maps since ASPP module guarantees the change of the receptive field. The feature maps in the front part tend to learn textures, and the feature maps in the back tend to learn edges. In this way our method can achieve better performance.

\section{Conclusion}
\label{sec:conclusion}

In this paper, we propose a lightweight model named GPU-Net, which can learn diverse feature maps and get better segmentation performance. We test our method on three dataset and visualize several feature maps to proof the effectiveness and efficiency of our method. Our plug-and-play GP-module can also be applied to existing segmentation methods to further improve their performance with fewer parameters and fewer FLOPs, which shed the light on the further research.

\bibliographystyle{IEEEbib}
\bibliography{strings,refs}

\end{document}